# Kahramanmaraş – Gaziantep, Türkiye Mw 7.8 Earthquake on February 6, 2023: Preliminary Report on Strong Ground Motion and Building Response Estimations

George Papazafeiropoulos[1*] and Vagelis Plevris[2]

[1] School of Civil Engineering, National Technical University of Athens
Zografou Campus, 15780 Athens, Greece
e-mail: gpapazafeiropoulos@yahoo.gr

[2] Department of Civil and Architectural Engineering, Qatar University
P.O. Box: 2713, Doha, Qatar
e-mail: vplevris@qu.edu.qa

*Corresponding author

**Abstract.** The effects on structures of the earthquake with magnitude 7.8 on the Richter scale (moment magnitude scale) which took place in Pazarcık, Kahramanmaraş, Türkiye at 04:17 a.m. local time (01:17 UTC) on February 6, 2023, are investigated by processing suitable seismic records using the open-source software OpenSeismoMatlab. The earthquake had a maximum Mercalli intensity of XI (Extreme) and it was followed by a Mw 7.5 earthquake nine hours later, centered 95 km to the north–northeast from the first. Peak and cumulative seismic measures as well as elastic response spectra, constant ductility (or isoductile) response spectra, and incremental dynamic analysis curves were calculated for two representative earthquake records of the main event. Furthermore, the acceleration response spectra of a large set of records were compared to the acceleration design spectrum of the Turkish seismic code. Based on the study, it is concluded that the structures were overloaded far beyond their normal design levels. This, in combination with considerable vertical seismic components, was a contributing factor towards the collapse of many buildings in the region. Modifications of the Turkish seismic code are required so that higher spectral acceleration values can be prescribed, especially in earthquake-prone regions.

## 1 Situation overview

An earthquake with magnitude (Mw) 7.8 took place in Pazarcık, Kahramanmaraş, Türkiye at 04:17 a.m. local time (01:17 UTC) on February 6, 2023. The earthquake had a maximum Mercalli intensity of XI (Extreme) and it was followed by a Mw 7.5 earthquake nine hours later, centered 95 km to the north–northeast from the first. According to information available as of February 22, 2023, and a press release of the Turkish government [1] 42,310 people have lost their lives in Kahramanmaraş, Gaziantep, Şanlıurfa, Diyarbakır, Adana, Adıyaman, Osmaniye, Hatay, Kilis and Malatya and Elazığ, and 448,010 people have been evacuated from the earthquake zone. A total of 7,184 aftershocks occurred and a total of 5,606 buildings have reportedly been destroyed in Türkiye [2]. In Türkiye and Syria combined, more than 6,500 buildings have collapsed due to the two main shocks. As of February 6, 2023, a three-month state of emergency is in place in provinces directly affected by the earthquake in Türkiye [3]. The details of the earthquake event (from now on referred to as the 7.8 $M_w$ event) are shown in Table 1 [4].

**Table 1**. Details of the main 7.8 $M_w$ earthquake event [4].

| | |
|---|---|
| **Magnitude** | 7.8 ($M_w$) |
| **Location** | Pazarcık (Kahramanmaraş), 26 km ENE of Nurdağı, Türkiye |
| **Date and time** | 6 February 2023, 01:17:34 UTC |
| **Latitude** | 37.225° N |
| **Longitude** | 37.021° E |
| **Depth** | 10.0 km |

## 2   Record data

Two representative seismic recording stations were selected for obtaining acceleration time history data of the 7.8 $M_w$ earthquake event: (i) the Station No 3137, TK Network, (Lat.: 36.69293°, Long.: 36.48852°), and (ii) the KHMN Station, KO Network KO, (Lat.: 37.3916°, Long.: 37.1574°), both shown in Figure 1, together with the epicenter of the earthquake event. For the processing of the acceleration time histories, the open-source Matlab code OpenSeismoMatlab [5] was used, which has been developed by the authors and is quite reliable since it has been successfully verified in several cases in the literature [6-8].

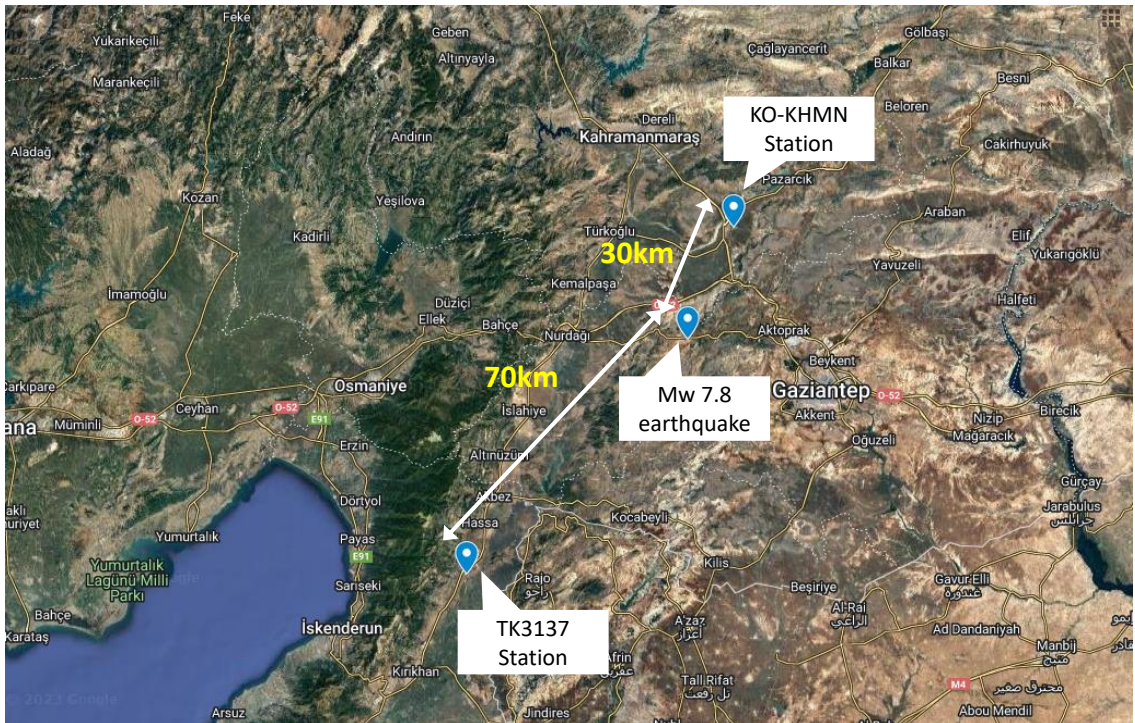

**Figure 1**. Locations of the Station No 3137, Network TK, and the KHMN Station, Network KO.

Figure 2 shows the two horizontal acceleration components of the TK3137 station record, while Figure 3 shows the corresponding vertical acceleration component. Similarly, Figure 3 shows the two horizontal acceleration components of the KO-KHMN station record, and Figure 4 shows its vertical acceleration component. The severity of the ground motions is obvious since the TK3137 record has reached a peak ground acceleration (PGA) of 0.75 g during ground shaking, whereas the KO-KHMN record has reached PGA values as high as 0.60 g. Another observation is that the duration of the shaking is quite large reaching about 1.5 minutes in the case of the TK3137 station.

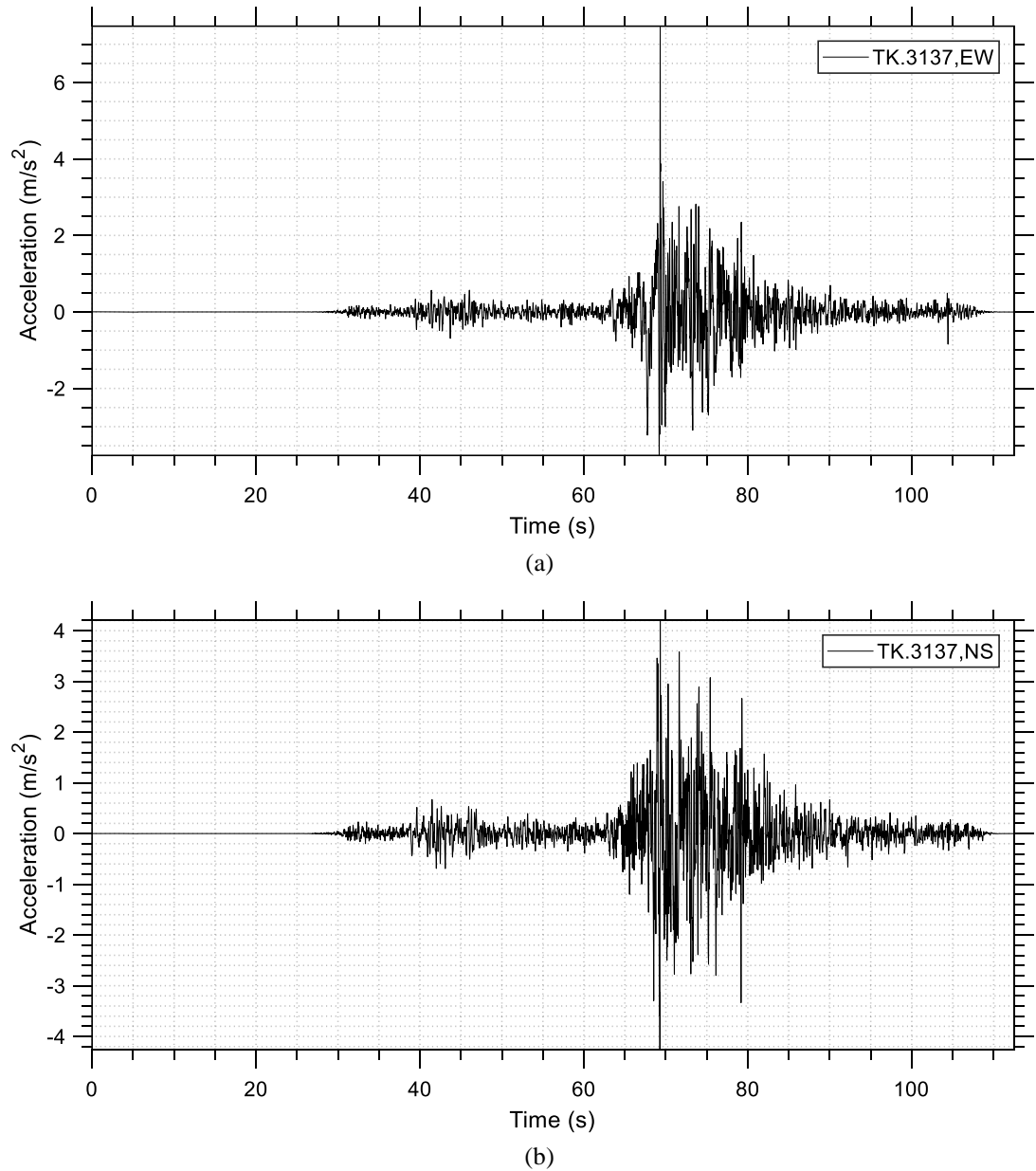

(a)

(b)

**Figure 2**. Earthquake acceleration time histories of the $M_w$ 7.8 earthquake recorded at the TK3137 station: (a) East-West component, (b) North-South component.

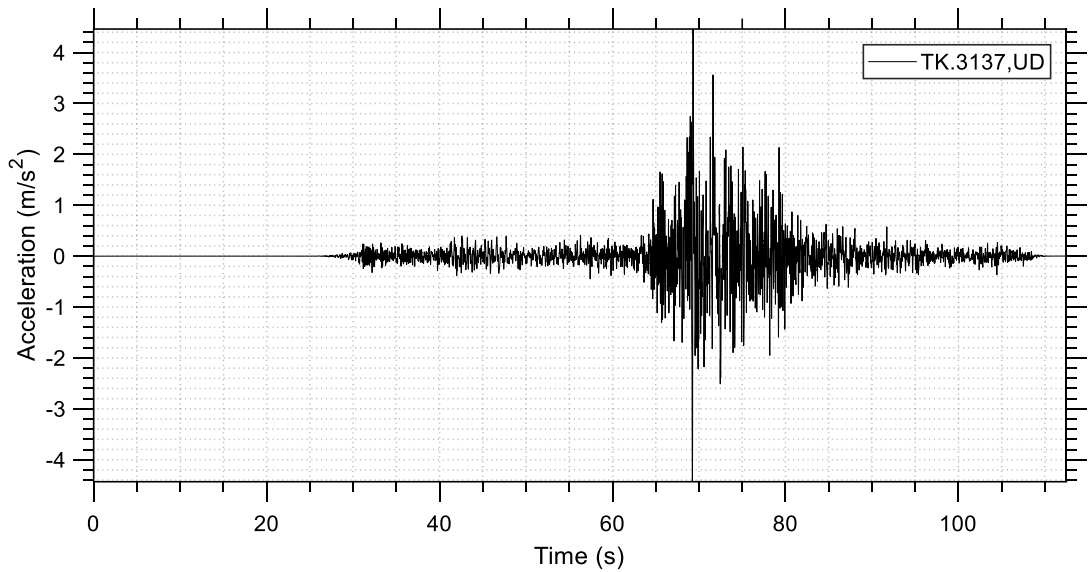

**Figure 3**. Earthquake acceleration time history of the $M_w$ 7.8 earthquake recorded at the TK3137 station: Vertical component.

Table 2 shows the peak values recorded for each station, namely the PGA as well as the peak ground velocity (PGV) and the peak ground displacement (PGD) for both records, for the horizontal (EW, NS) and the vertical (UD) components.

**Table 2**. Peak seismic parameters of the 7.8 $M_w$ earthquake even, based on the two recordings.

| Station | TK3137 | | | KO-KHMN | | |
|---|---|---|---|---|---|---|
| Component | PGA (m/s²) | PGV (m/s) | PGD (m) | PGA (m/s²) | PGV (m/s) | PGD (m) |
| EW Horizontal | 7.47 | 0.75 | 0.50 | 5.09 | 1.08 | 0.61 |
| NS Horizontal | 4.26 | 0.76 | 1.15 | 6.06 | 0.89 | 0.50 |
| UD Vertical | 4.46 | 0.40 | 0.16 | 4.79 | 0.45 | 0.34 |

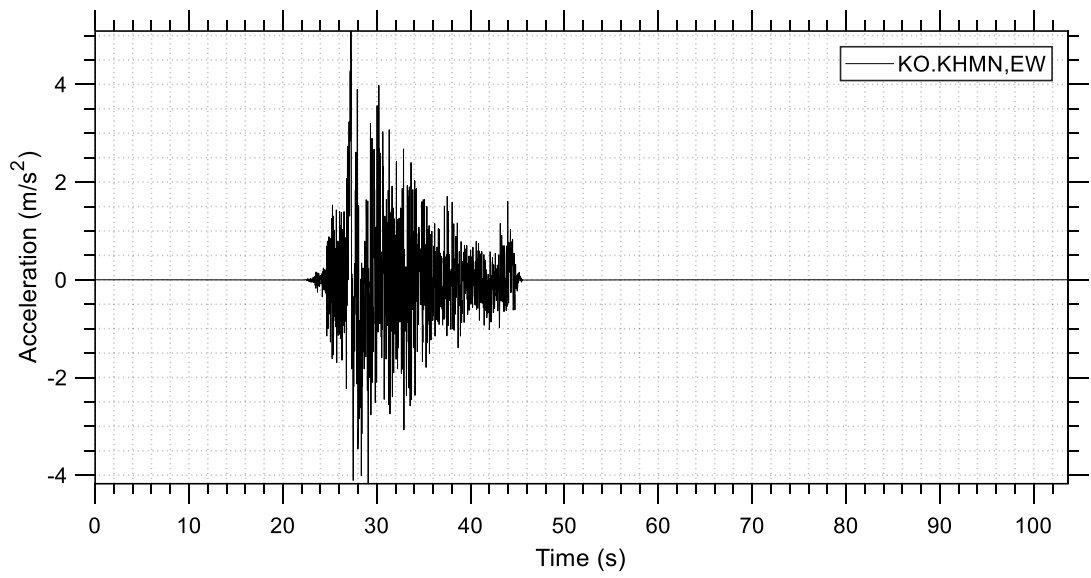

(a)

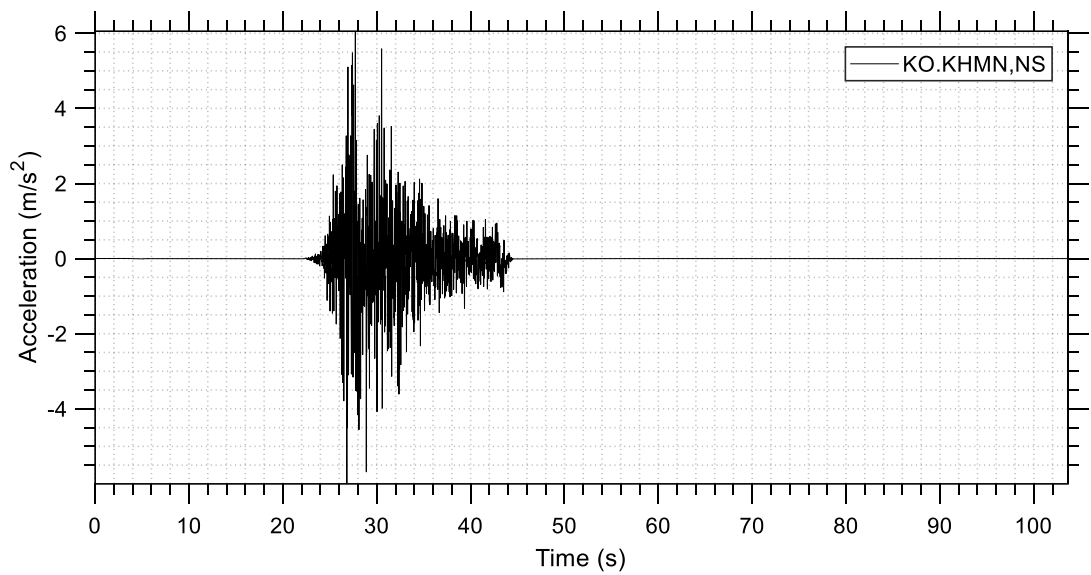

(b)

**Figure 4**. Earthquake acceleration time histories of the $M_w$ 7.8 earthquake recorded at the KHMN Station: (a) East-West component, (b) North-South component.

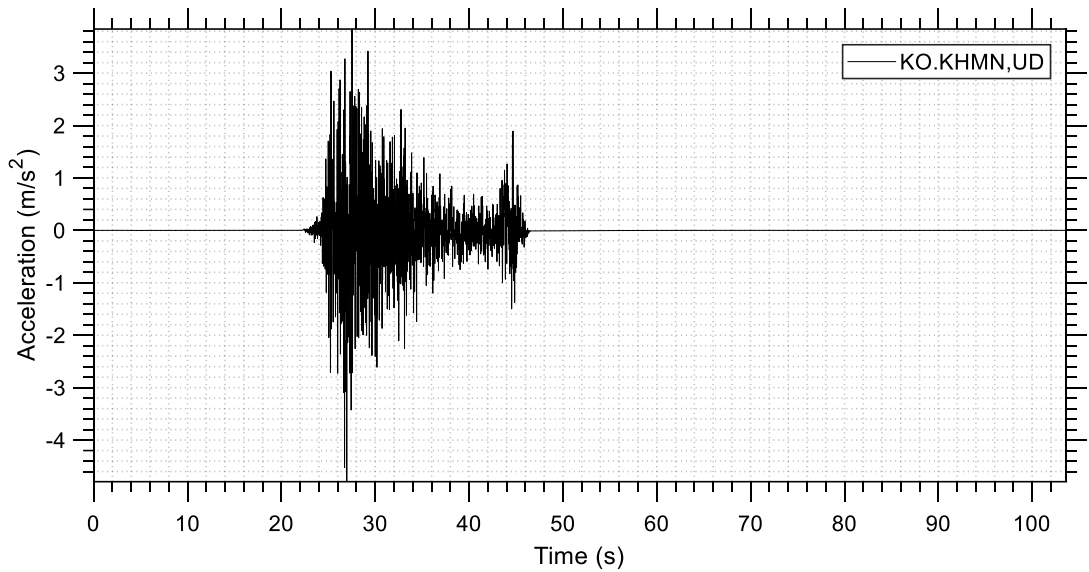

**Figure 5**. Earthquake acceleration time histories of the $M_w$ 7.8 earthquake recorded at the KHMN Station: Vertical component.

## 2.1 Cumulative energy, Arias intensity, and significant duration data

A significant measure of the destructive effect of an earthquake on structures is the amount of energy that it releases. This energy, after having been released by the earthquake is partly absorbed by the structures in the area and this results in their gradual damage, or even collapse. Energy-based design theory (EBDT), which introduces energy demand as the critical parameter to evaluate structural damage, has gained attention around the world in recent decades. According to this approach, each structure must be designed to be capable of absorbing a certain amount of energy. It is noted that the EBDT concept accounts for the cumulative damage and the effective duration of earthquakes in a theoretically sound manner. The effects of the latter factors are not taken sufficiently into account in traditional force-based design approaches. In this section, the data for the total cumulative energy, Arias intensity and significant duration are presented for the two records.

A summary report is given in Table 3 where *Ecum* denotes the cumulative energy, *Arias* denotes the Arias intensity, $td_{5-95}$ denotes the time needed for the 90% of the seismic energy to be released (5%-95% interval), and $td_{5-75}$ denotes the time needed for the 70% of the seismic energy to be released (5%-75% interval). The main trend, apart from the relatively high Arias intensity and cumulative energy values, is that the seismic energy was released in a small fraction of the total duration of the earthquake. For example, for station TK3137, it took only 16 sec for 90% of the seismic energy to be released, of which only roughly 8 sec to release 70% of it. If these time durations are compared to the total duration of the earthquake, which is larger than 80 seconds, it becomes apparent that the $M_w$ 7.8 event was an event of large seismic power. This fact played a critical role in the intensity of the shaking that was experienced by structures and could provide some indirect hints explaining the large number of structural collapses. The aforementioned points become obvious by observing the normalized cumulative energy time histories shown in Figure 6.

**Table 3**. Cumulative seismic parameters of the 7.8 $M_w$ earthquake event.

| Station | TK3137 | | | | KO-KHMN | | | |
|---------|--------|--------|--------|-------|---------|--------|--------|-------|
| Component | Ecum (m²/sec³) | td$_{5-95}$ (sec) | td$_{5-75}$ (sec) | Arias (m/s) | Ecum (m²/sec³) | td$_{5-95}$ (sec) | td$_{5-75}$ (sec) | Arias (m/s) |
| EW | 22.755 | 16.27 | 8.26 | 3.6 | 21.220 | 12.145 | 5.88 | 3.4 |
| NS | 22.232 | 16.79 | 9.58 | 3.6 | 28.743 | 10.28 | 4.64 | 4.6 |
| UD | 13.932 | 16.68 | 9.71 | 2.2 | 13.731 | 16.59 | 5.44 | 2.2 |

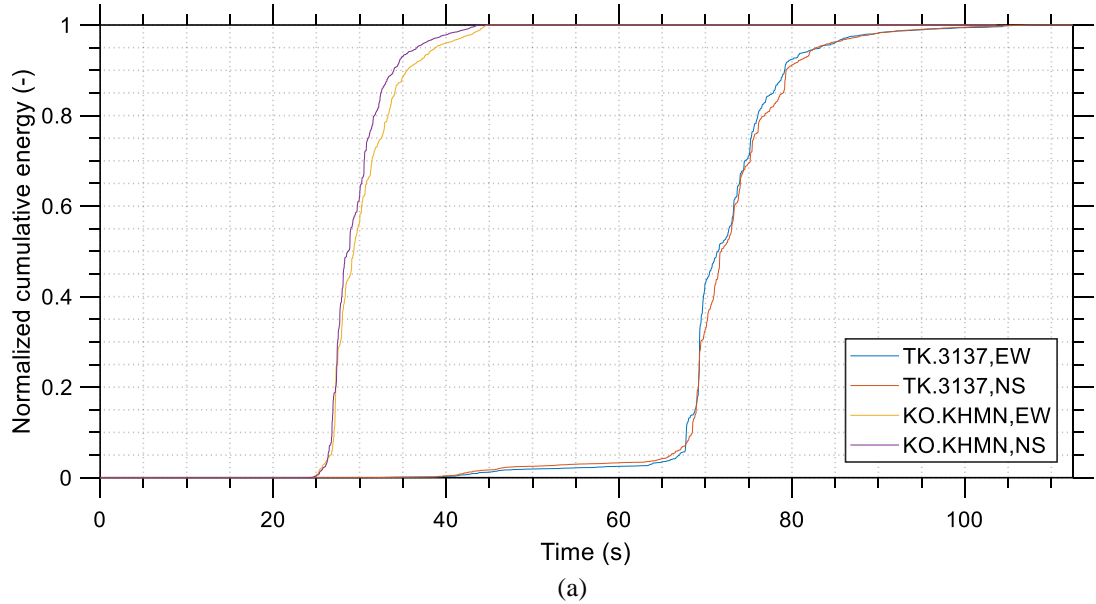

(a)

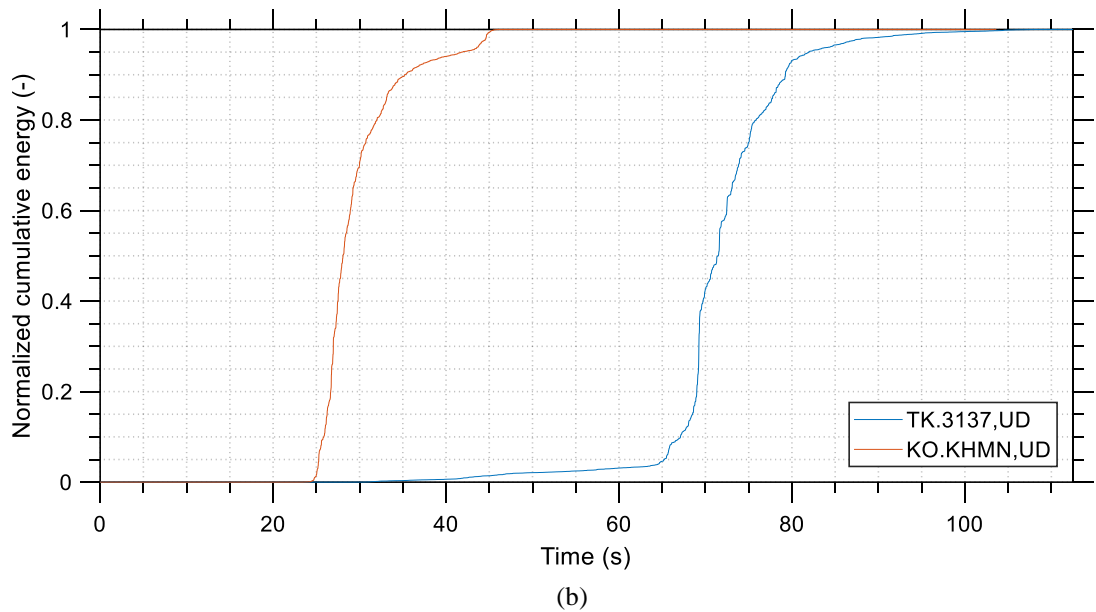

(b)

**Figure 6**. Normalized cumulative energy that was released by the $M_w$ 7.8 earthquake recorded, as recorded at the TK3137 and the KO-KHMN stations: (a) Horizontal components, (b) Vertical components.

## 2.2 Elastic response spectra

A response spectrum is a plot that shows the maximum response of a structure to a ground motion as a function of frequency (or equivalently, period). In structural analysis, response spectrum analysis is a method used to evaluate the dynamic response of a structure to earthquake loads. The response spectrum is derived from a set of ground motion records that represent the expected seismic hazard for the region where the structure is located. These ground motion records are typically generated by seismologists and are based on historical seismic data, geological surveys, and other sources of information. Response spectra are used in response spectrum modal analysis (RSMA), a method used to analyze the dynamic behavior of structures subjected to earthquake loads, by determining the maximum response of a structure to earthquake ground motion. RSMA allows engineers to design structures that can withstand the expected earthquake loads in a region. By analyzing the structure's response to different ground motion records, given the uncertainties involved, the engineers can hope that the structure will not experience excessive deformation or failure during an earthquake. This analysis is commonly used in the design of buildings, bridges, and other structures that are critical to public safety.

The elastic response spectra simulate a single degree of freedom (SDOF) system and the way that it would respond to a given earthquake time history. We are interested both in the peak and the cumulative spectral response of a SDOF system for an earthquake. Typical spectral quantities of the first category are spectral acceleration and pseudoacceleration, spectral velocity and pseudovelocity, and spectral displacement. Typical spectral quantities of the latter category are the seismic input energy equivalent velocity [9], absolute and relative. It is noted that the seismic input energy equivalent velocity is a measure of the seismic energy that is input to a structure by an earthquake, and not the energy that is released by the earthquake itself. Depending on the dynamic properties of the structure, the amount of seismic energy that it absorbs is different.

The spectral displacement, velocity, and acceleration are shown in Figure 7, Figure 8, and Figure 9, respectively, for the two records and their different components (the two horizontal and the vertical). Similarly, Figure 10 and Figure 11 show the spectral pseudoacceleration and the spectral pseudovelocity, respectively. All diagrams have been generated for damping ratio ζ equal to 5% and their horizontal axis is in logarithmic scale. In Figure 9, it can be observed that the NS component of the KO-KHMN station record reached very high spectral accelerations, around 2.35 g, whereas the vertical component of the same record reached a maximum spectral acceleration close to 1.7 g. These spectral acceleration values are extraordinarily high. On the other hand, it is seen from Figure 8 that at the high period range the spectral velocity of the record of the TK3137 station is generally larger (with some exceptions), for both its horizontal and vertical components. This is an important observation that may explain the large casualties that occurred in the Hatay region, although it is far away from the epicenter of the main earthquake, especially in comparison to Kahramanmaraş. The TK3137 station, which gave higher spectral velocities than the KO-KHMN station, is much closer to the Hatay region, as shown in Figure 1. This may imply a stronger relationship between the destructive effects of an earthquake and its spectral velocity, rather than its spectral acceleration. While the spectral acceleration has been traditionally taken into account for the design of buildings according to various seismic norms worldwide (including the Turkish seismic code), the spectral velocity is generally ignored although it may be a better index in determining seismic hazard for taller buildings and it can serve as a parameter from which to estimate the macroseismic intensity and structural damage [10]. This is a well-known issue, and the $M_w$ 7.8 earthquake may provide an incentive for further improvements to the seismic codes in this direction.

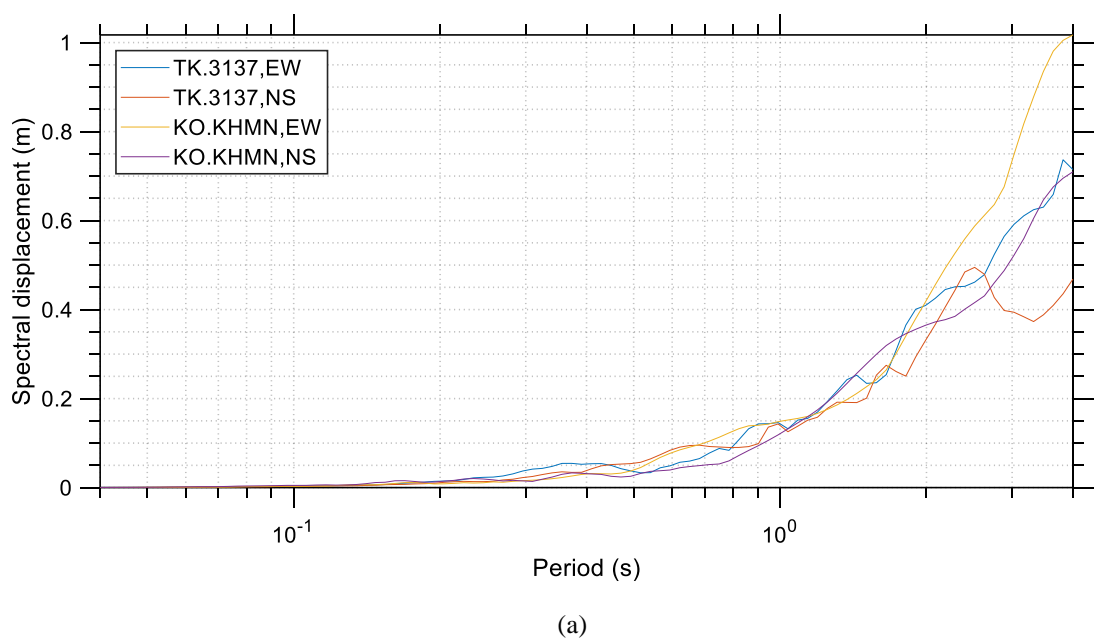

(a)

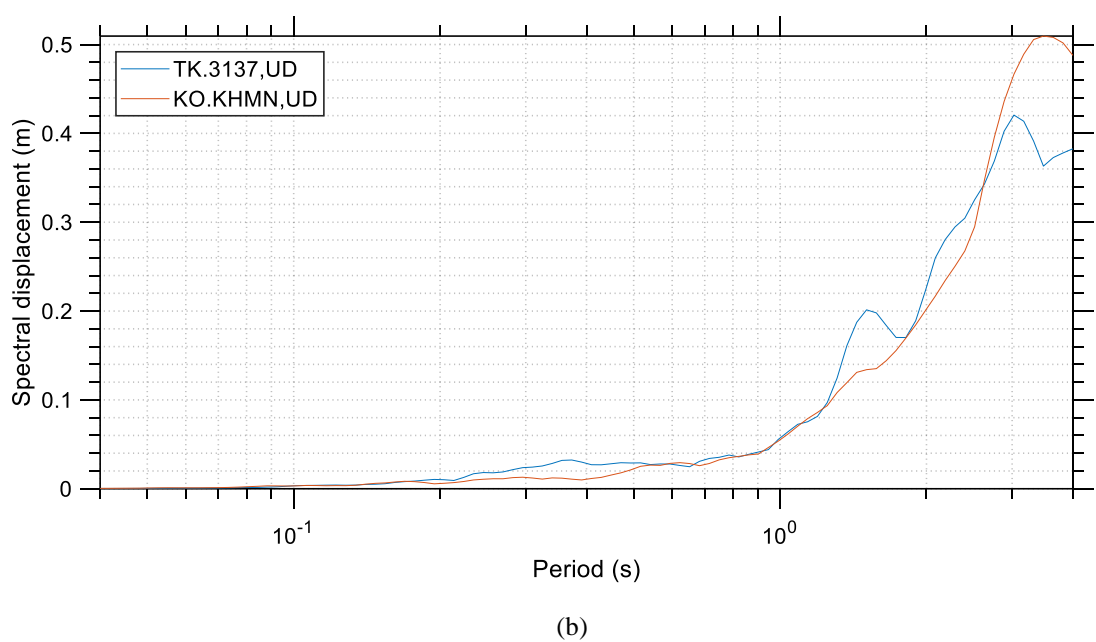

(b)

**Figure 7**. Spectral displacement ($\zeta$=5%) for the $M_w$ 7.8 earthquake, as recorded at the TK3137 and the KO-KHMN stations: (a) Horizontal components, (b) Vertical components.

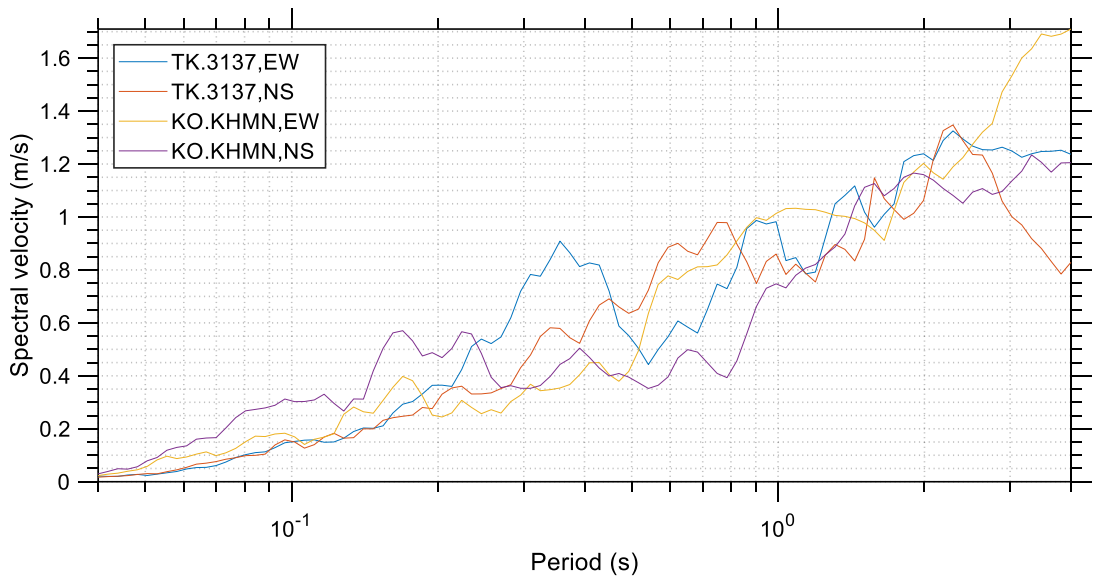

(a)

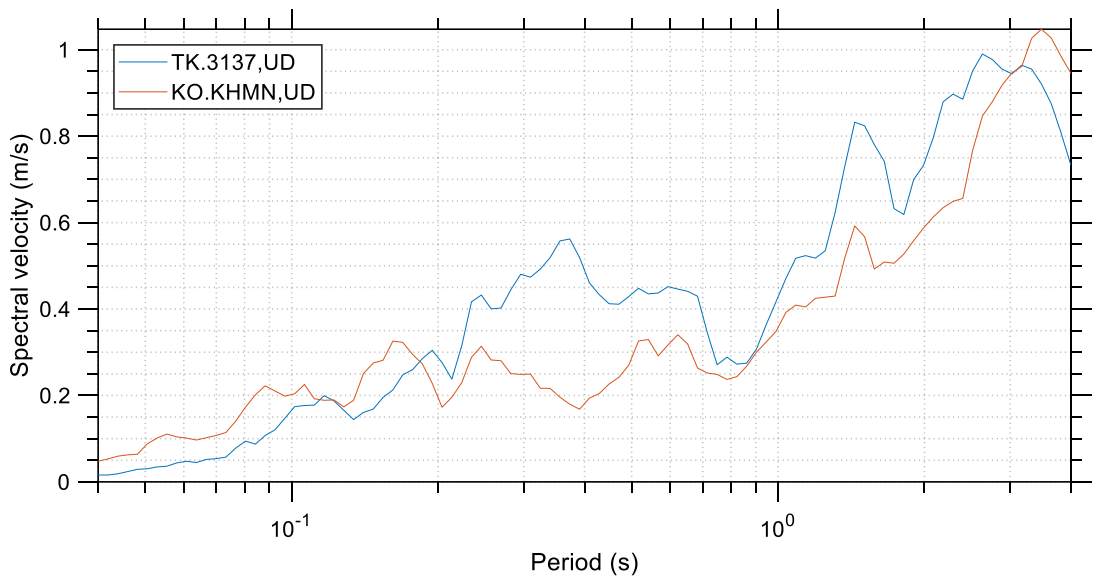

(b)

**Figure 8**. Spectral velocity ($\zeta$=5%) for the $M_w$ 7.8 earthquake, as recorded at the TK3137 and the KO-KHMN stations: (a) Horizontal components, (b) Vertical components.

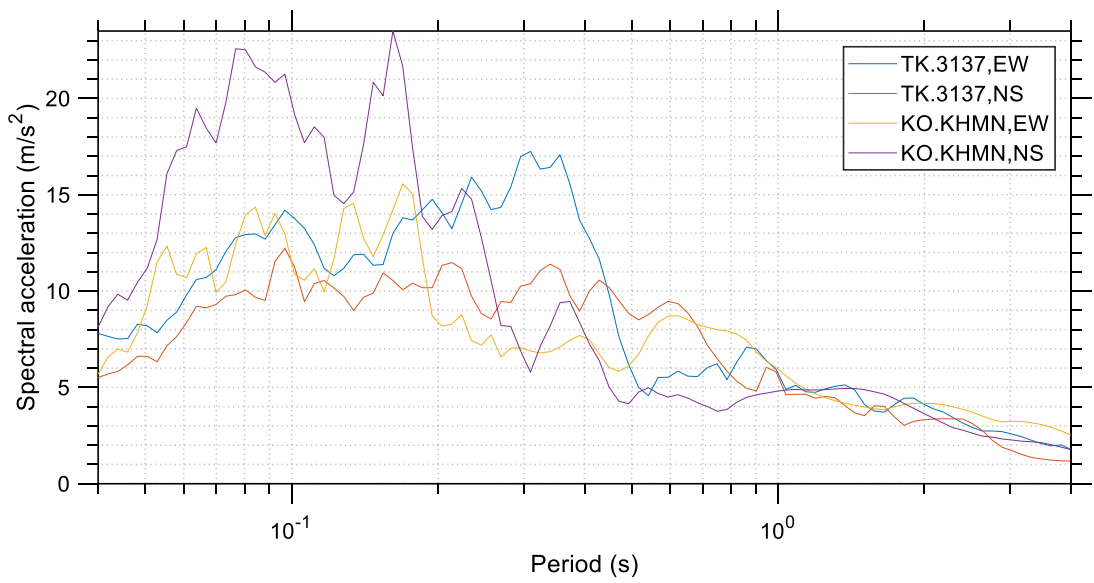

(a)

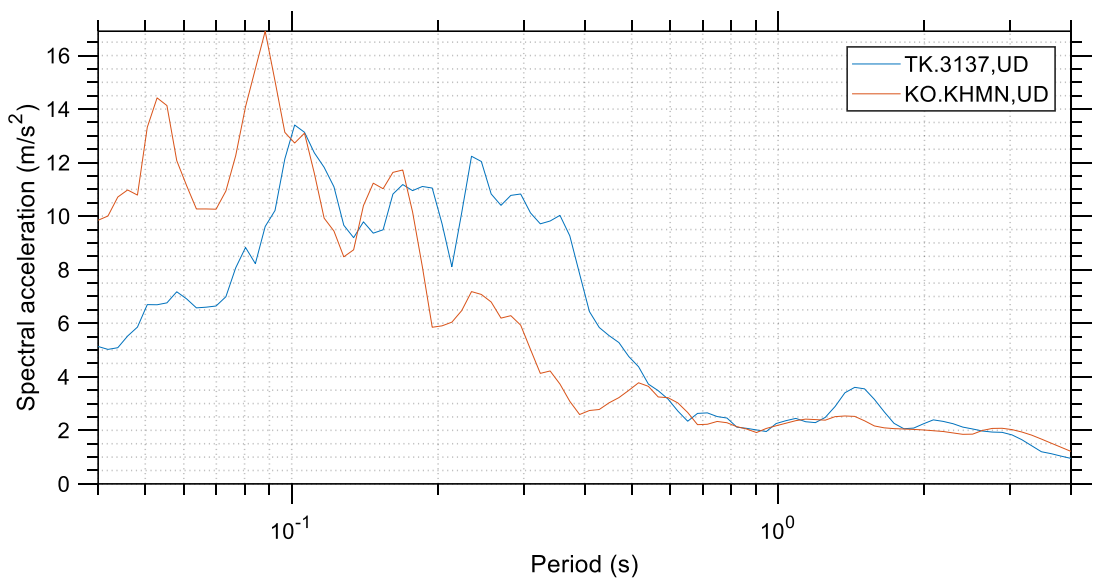

(b)

**Figure 9**. Spectral acceleration ($\zeta$=5%) for the $M_w$ 7.8 earthquake, as recorded at the TK3137 and the KO-KHMN stations: (a) Horizontal components, (b) Vertical components.

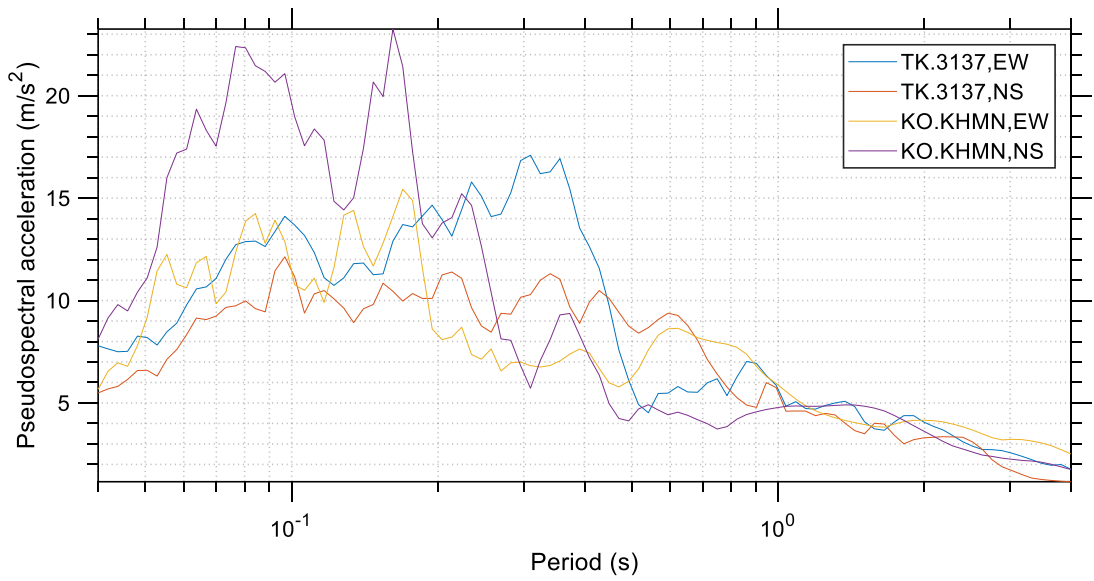

(a)

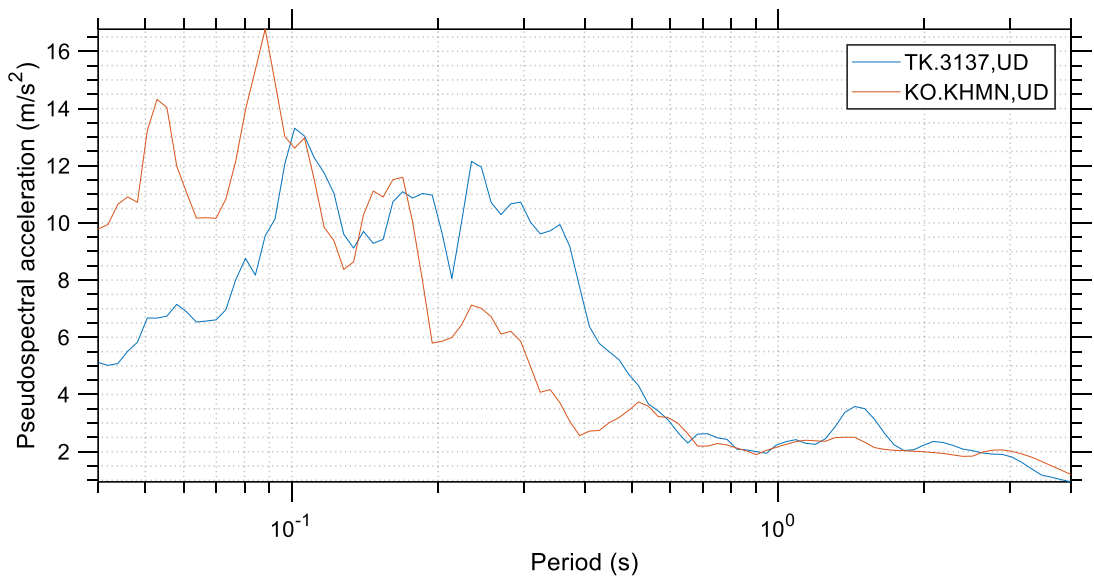

(b)

**Figure 10**. Spectral pseudoacceleration ($\zeta$=5%) for the $M_w$ 7.8 earthquake, as recorded at the TK3137 and the KO-KHMN stations: (a) Horizontal components, (b) Vertical components.

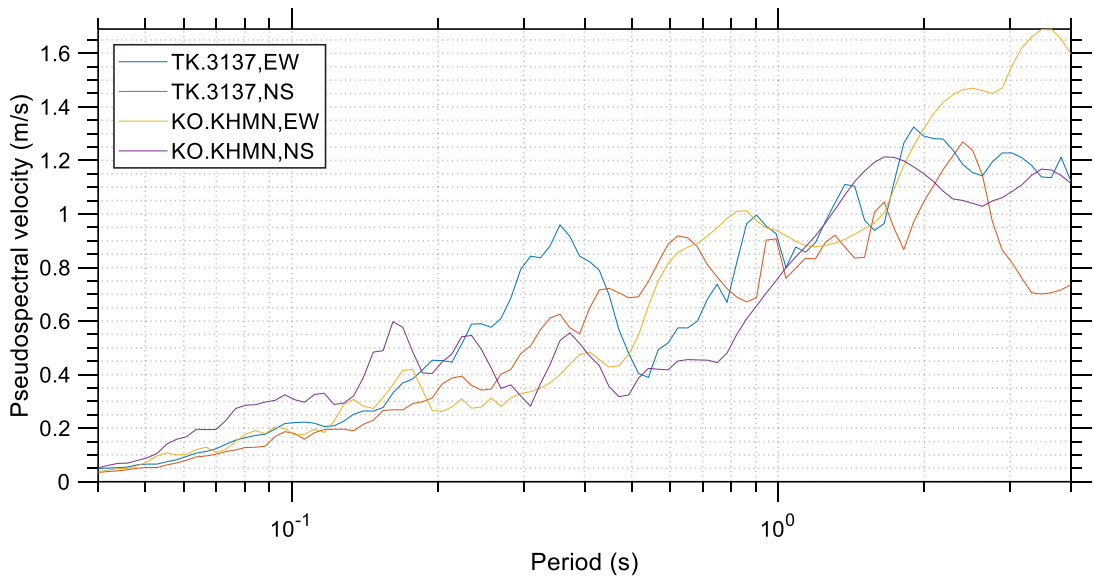

(a)

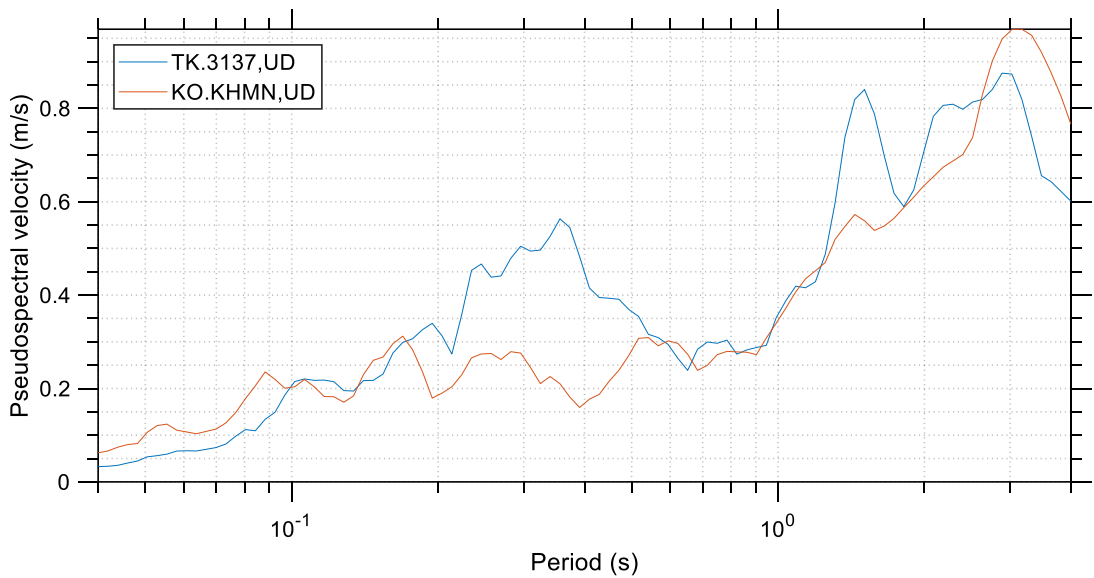

(b)

**Figure 11**. Spectral pseudovelocity ($\zeta$=5%) for the $M_w$ 7.8 earthquake, as recorded at the TK3137 and the KO-KHMN stations: (a) Horizontal components, (b) Vertical components.

## 2.3 Isoductile response spectra

Seismic energy can be absorbed in structures in two main forms: (i) as elastic strain energy and (i) as hysteretically dissipated energy. The former approach requires that the structure remains in the elastic region and resists the entire earthquake load, as high as its peak value, elastically. The latter approach permits the design of structures based on reduced earthquake loads (lower than those of the former case, which correspond to the maximum acceleration that the structure experiences during the seismic event) by relying on ductility and over-strength of the materials used. A structure that responds in an elastoplastic way through its ductile character experiences lower acceleration during an earthquake and its design becomes more economical, although damages may be expected in case of high accelerations. The basic ductile design philosophy is that the structure should survive the main shock through controlled damage but without collapse.

The constant ductility (or isoductile) spectra assume that a SDOF system responds in an elastoplastic (i.e., elastic – perfectly plastic) way with constant ductility, which is defined as the ratio of the maximum displacement to the yielding displacement. The yielding displacement is the displacement that corresponds to the yield limit of the SDOF system. In other words, we are interested in the response of SDOF systems for varying eigenperiods, similar to the elastic spectra considered in the previous section, but for a specific ductility ratio. The isoductile spectral displacement, velocity, and acceleration are shown in Figure 12, Figure 13, and Figure 14, respectively, for the two records and their different components (the two horizontal and the vertical). Similarly, Figure 15 and Figure 16 show the isoductile spectral pseudoacceleration and spectral pseudovelocity, respectively. All diagrams correspond to damping ratio $\zeta$ equal to 5%, ductility $\mu$ equal to 2 and their horizontal axis is in logarithmic scale. As expected, the comparison between the linear elastic and the isoductile spectra reveals that the maximum responses in the isoductile spectra are generally lower than those in the elastic spectra. For example, the maximum spectral acceleration of the horizontal components of the $M_w$ 7.8 event is equal to 2.35 g as shown in Figure 9 whereas the corresponding value for the isoductile spectra is equal to 1.4 g as shown in Figure 14. The difference in the maximum spectral acceleration implies a substantial difference in the applied seismic forces, and this shows the importance of structural ductility. The collapses due to the $M_w$ 7.8 earthquake showed in many cases a nonductile, brittle behavior, which in the case of reinforced concrete (RC) structures is closely related to under-reinforced structural elements. These structures, having limited ductility, responded in a more linear elastic-wise manner, and thus experienced much larger accelerations, which explains many of the building collapses.

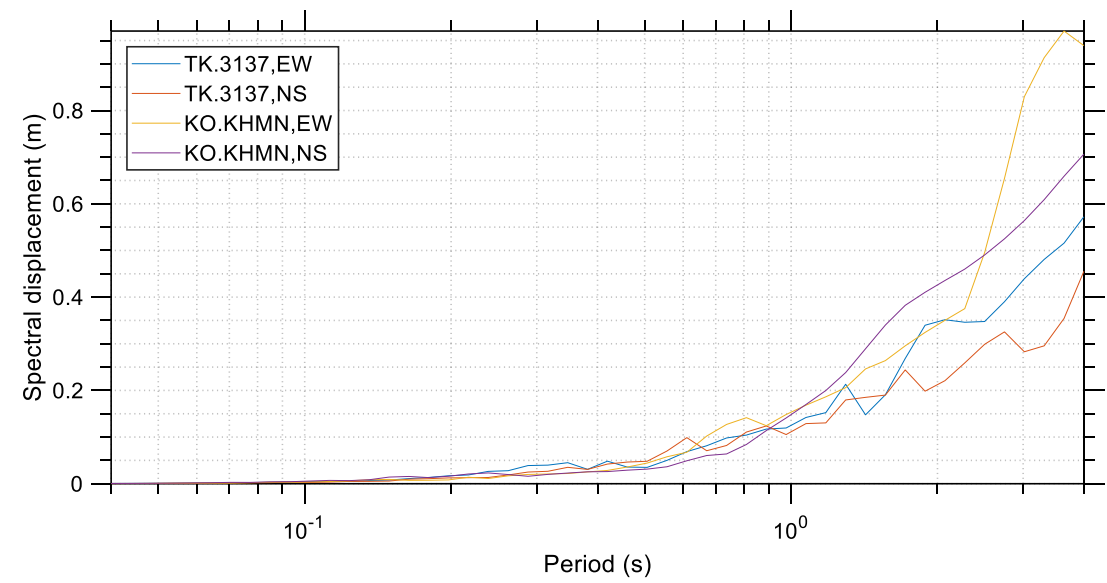

(a)

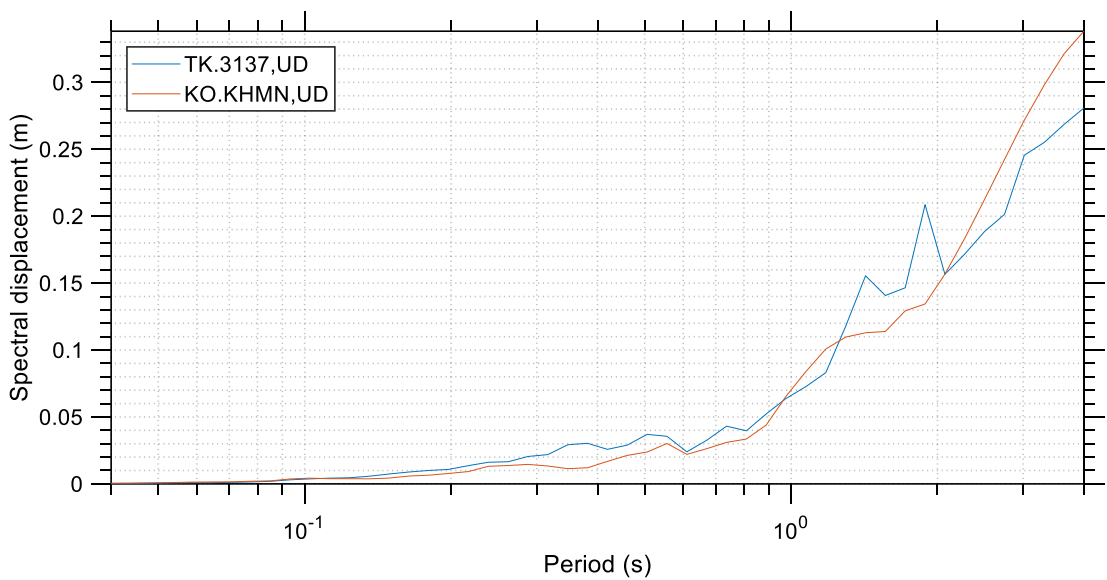

(b)

**Figure 12**. Isoductile spectral displacement ($\zeta$=5%, $\mu$=2) for the $M_w$ 7.8 earthquake, as recorded at the TK3137 and the KO-KHMN stations: (a) Horizontal components, (b) Vertical components.

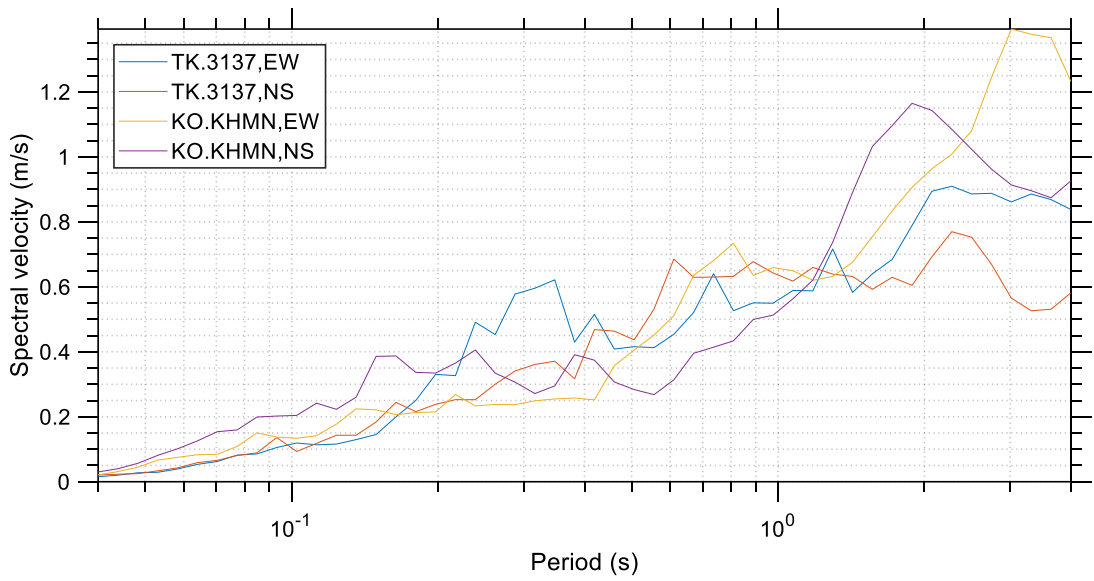

(a)

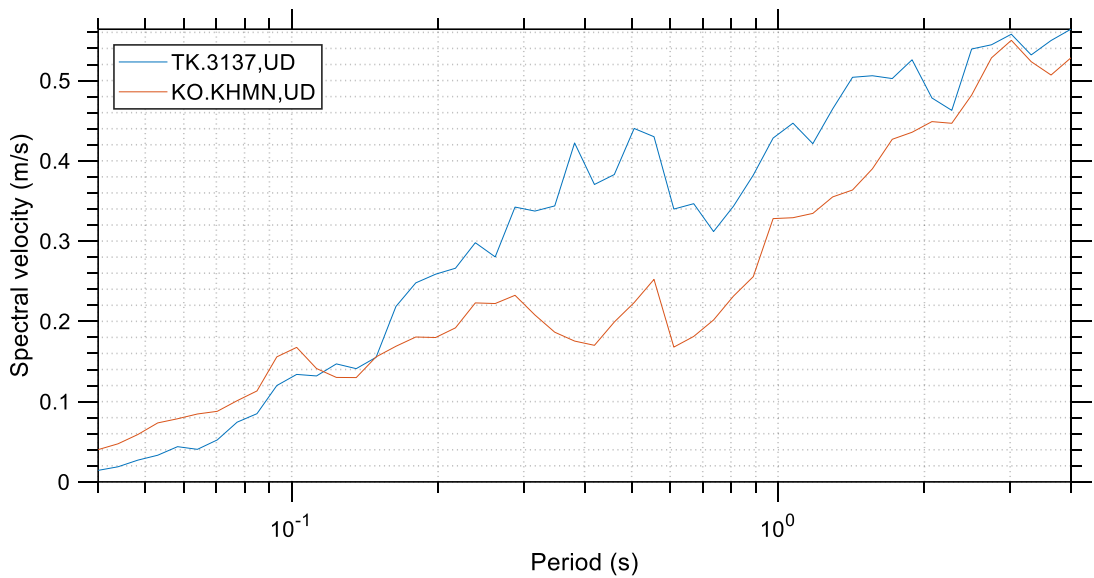

(b)

**Figure 13**. Isoductile spectral velocity ($\zeta$=5%, $\mu$=2) for the $M_w$ 7.8 earthquake, as recorded at the TK3137 and the KO-KHMN stations: (a) Horizontal components, (b) Vertical components.

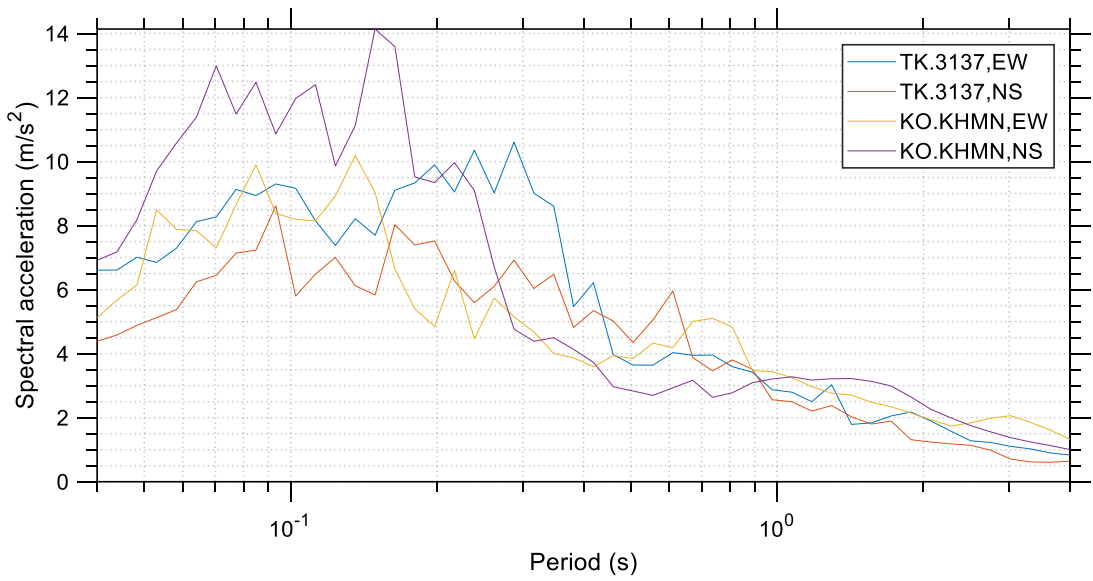

(a)

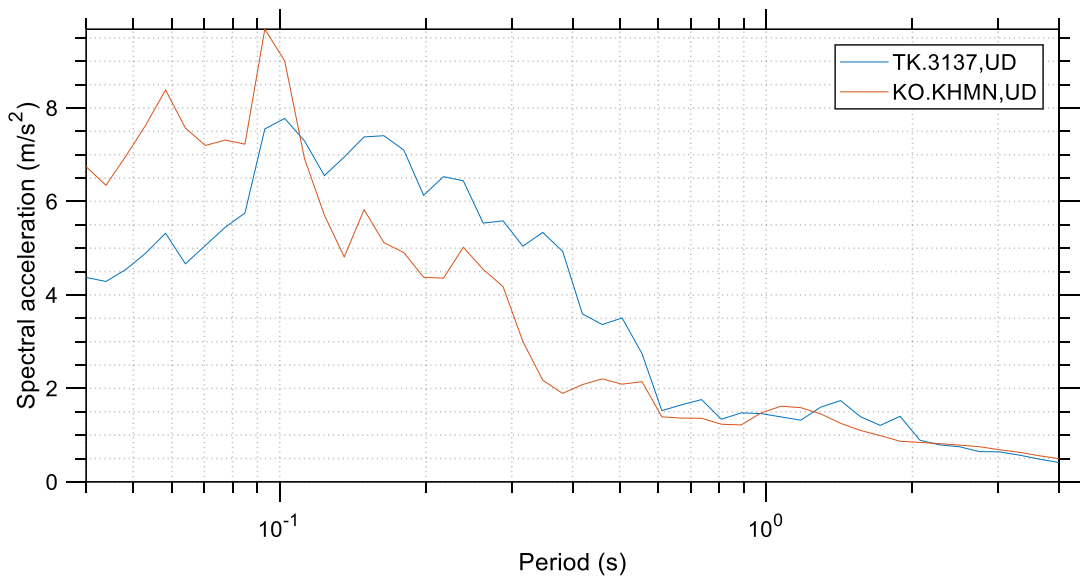

(b)

**Figure 14**. Isoductile spectral acceleration ($\zeta$=5%, $\mu$=2) for the M$_w$ 7.8 earthquake, as recorded at the TK3137 and the KO-KHMN stations: (a) Horizontal components, (b) Vertical components.

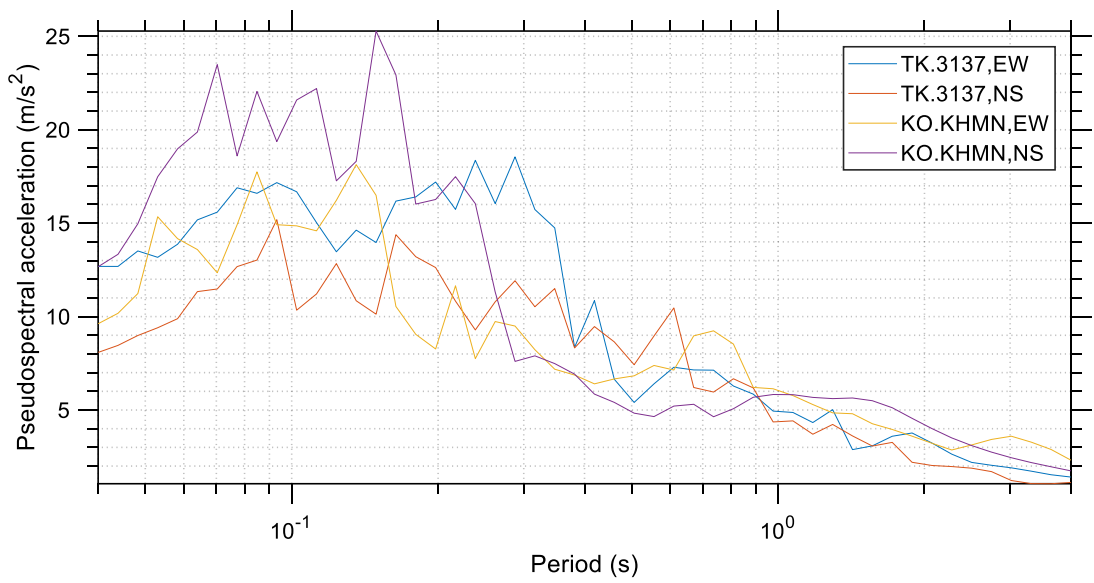

(a)

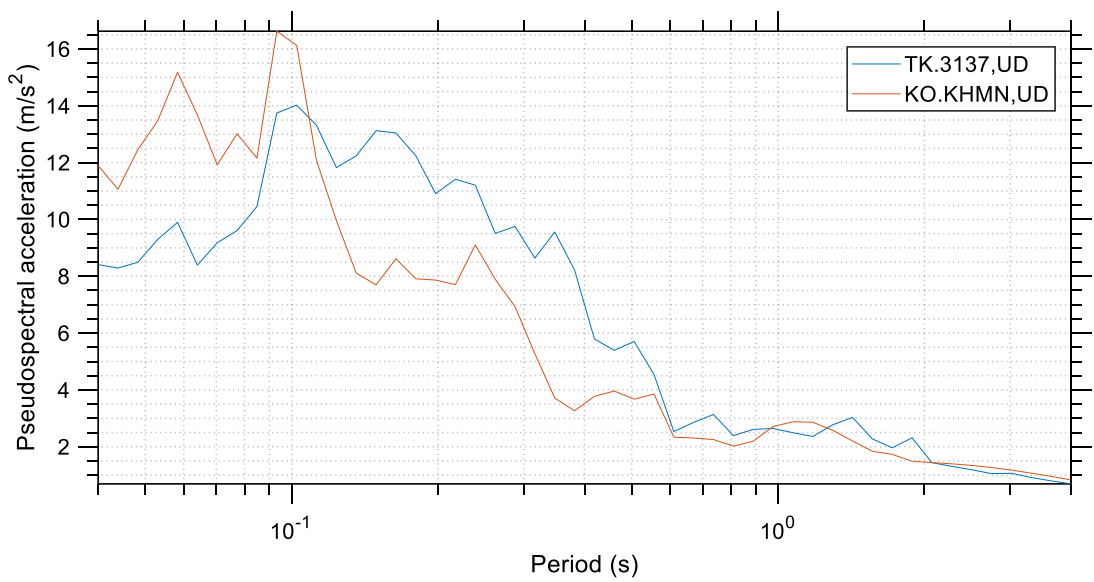

(b)

**Figure 15**. Isoductile spectral pseudoacceleration ($\zeta$=5%, $\mu$=2) for the $M_w$ 7.8 earthquake, as recorded at the TK3137 and the KO-KHMN stations: (a) Horizontal components, (b) Vertical components.

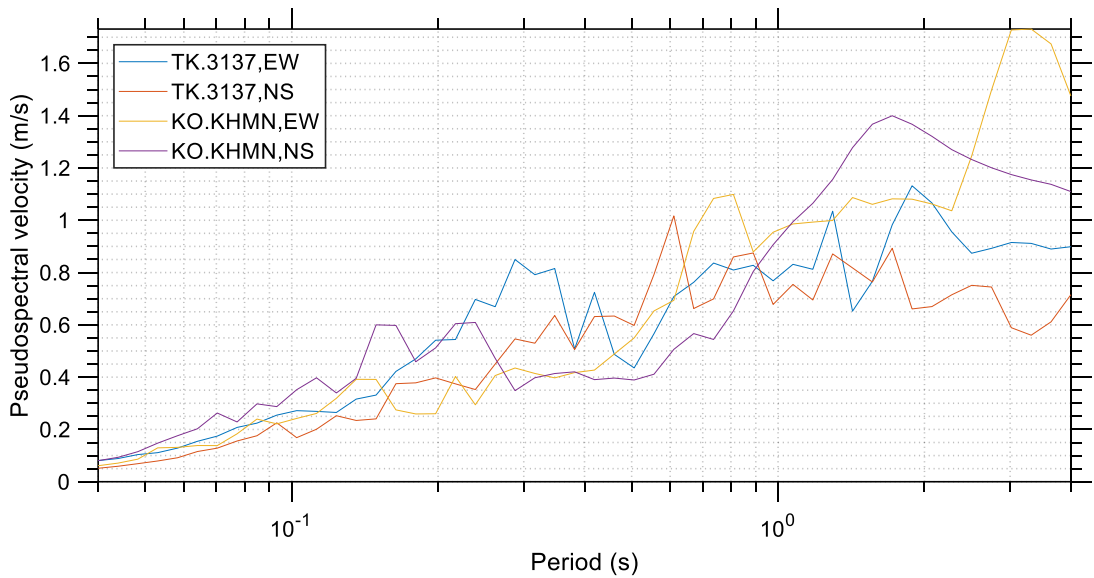

(a)

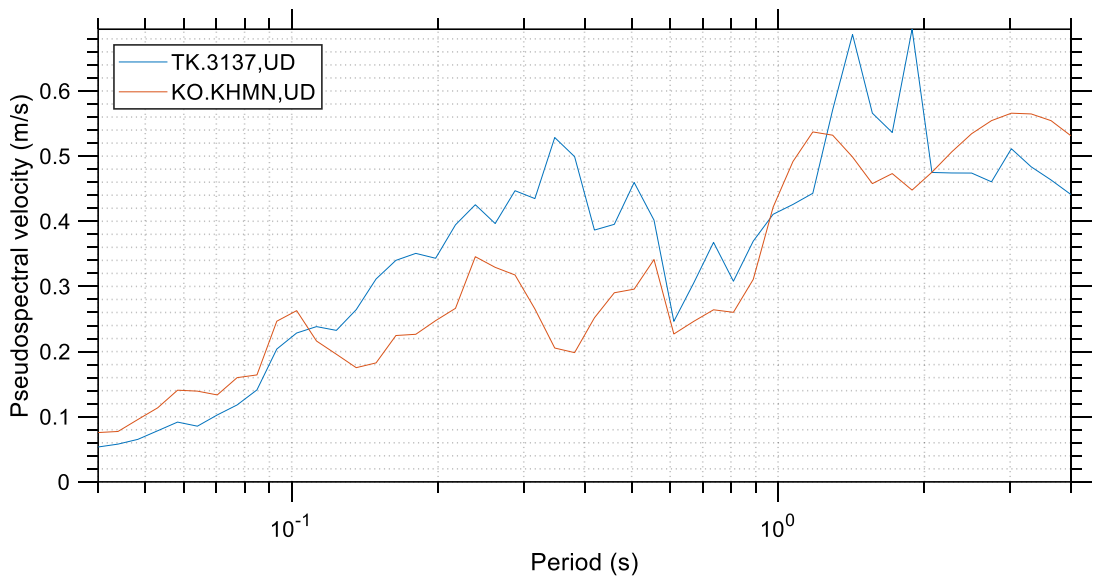

(b)

**Figure 16**. Isoductile spectral pseudovelocity ($\zeta$=5%, $\mu$=2) for the M$_w$ 7.8 earthquake, as recorded at the TK3137 and the KO-KHMN stations: (a) Horizontal components, (b) Vertical components.

## 2.4 The importance of ductility in structures

An important observation can be made based on the elastic and isoductile seismic response spectra: the maximum acceleration which is experienced by the elastoplastic ductile structures is substantially lower than that experienced by elastic structures. For example, in Figure 14, the maximum isoductile spectral acceleration for ductility $\mu$=2 is roughly 1.4 g and 0.97 g for the horizontal and vertical components, respectively. The corresponding values for linear elastic nonyielding structures can be seen in Figure 9, which are 2.35 g and 1.7 g for the horizontal and vertical components, respectively. In the horizontal direction, an increase from 1.4 g to 2.35 g accounts for 68% higher horizontal acceleration values for the nonductile structures. Low accelerations are directly related to low seismic forces, through Newton's second law, and thus lower requirements on behalf of the structure to resist these forces. Therefore, it becomes evident that ductility is an important aspect of seismic design since reduced seismic loads result in more economical designs. Apart from this, a structure of increased ductility is generally safer, since it can accommodate large deformations which cannot go unnoticed by the occupants and act as a warning for the imminent failure of the structure. This can save some critical time in difficult situations for the occupants when evacuation is required and potentially save their lives.

## 2.5 What does the Turkish Seismic code provide?

It is interesting to compare the effect of the earthquake event on the structures with the requirements of the Turkish Seismic code in the region. The comparison is made with reference to the records of the TK3137 and the KO-KHMN stations and it is shown in Figure 17 for the cases of linear elastic response spectra. Based on the comparison, the major conclusion is that the earthquake struck mainly at the low period range, where the design acceleration is 1.4 g and the maximum acceleration observed is equal to roughly 2.4 g. This is a significant difference, not only in the acceleration magnitude but also in its period content. Even for site class I, which contains the lowest period content (i.e. corresponds to stiff rock), there was significant acceleration below the lowest reference period. Therefore, suitable adjustments need to be made in the design spectrum of the Turkish code so that rare events, such as the one ($M_w$ 7.8) considered in this study, can be taken into account.

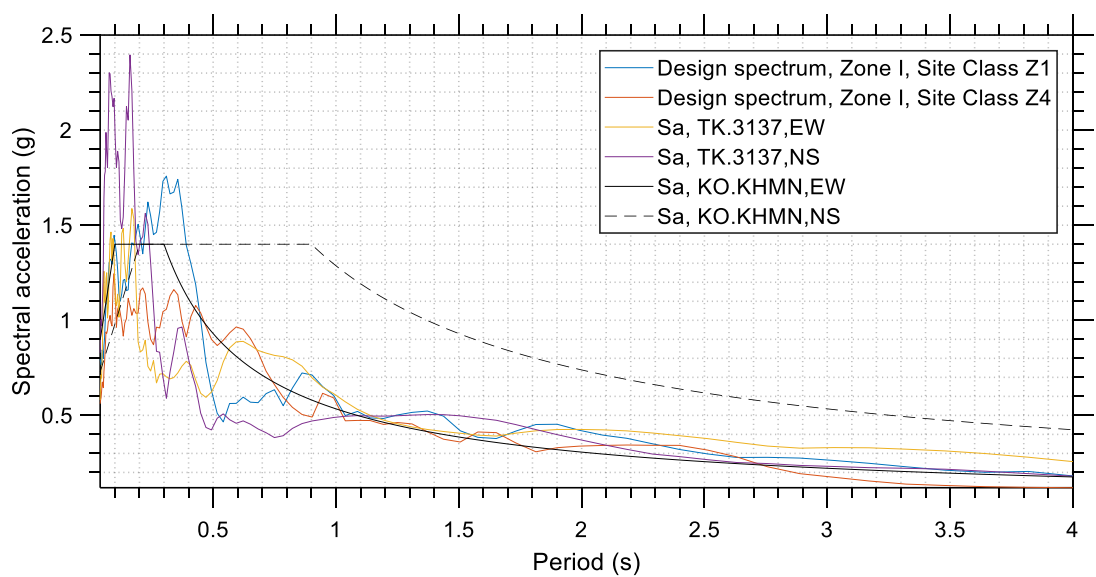

(a)

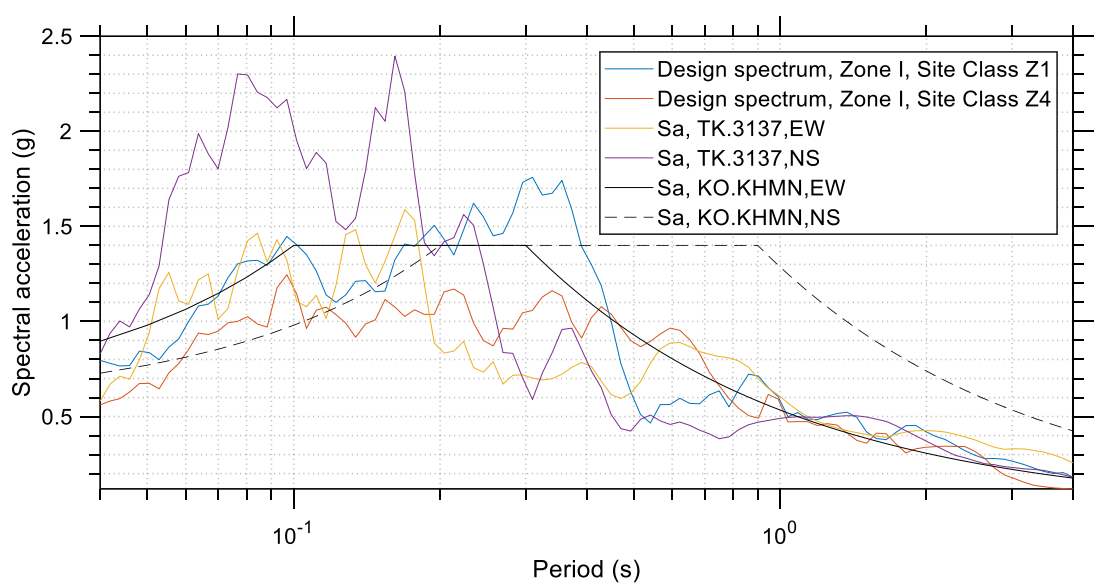

(b)

**Figure 17**. Design spectrum of the Turkish Seismic code vs actual acceleration response spectra for the $M_w$ 7.8 earthquake ($\zeta$=5%), as recorded at TK3137 and KO-KHMN stations: (a) Linear scale, (b) Logarithmic scale.

## 2.6 Does the high spectral acceleration in the low period range occur only for the two examined records or is it a general trend?

At this point, we need to check whether the trend that appears in Figure 17 is a general trend or it is specific only to these two recordings. For this purpose, more earthquake acceleration records need to be taken into account. In Figure 18 the acceleration spectra of several earthquake records are shown and compared to the provisions of the Turkish seismic code (linear elastic response spectra). It is shown that there are higher spectral acceleration values for a broader range of eigenperiods, for many of the recordings. Based on the envelope spectrum, a maximum spectral acceleration of 5.35 g is observed, which is extremely high, and is responsible for the many collapses due to the earthquake event. A need for a revision of the seismic code standards seems to exist, i.e., higher acceleration values for the design spectra must be proposed.

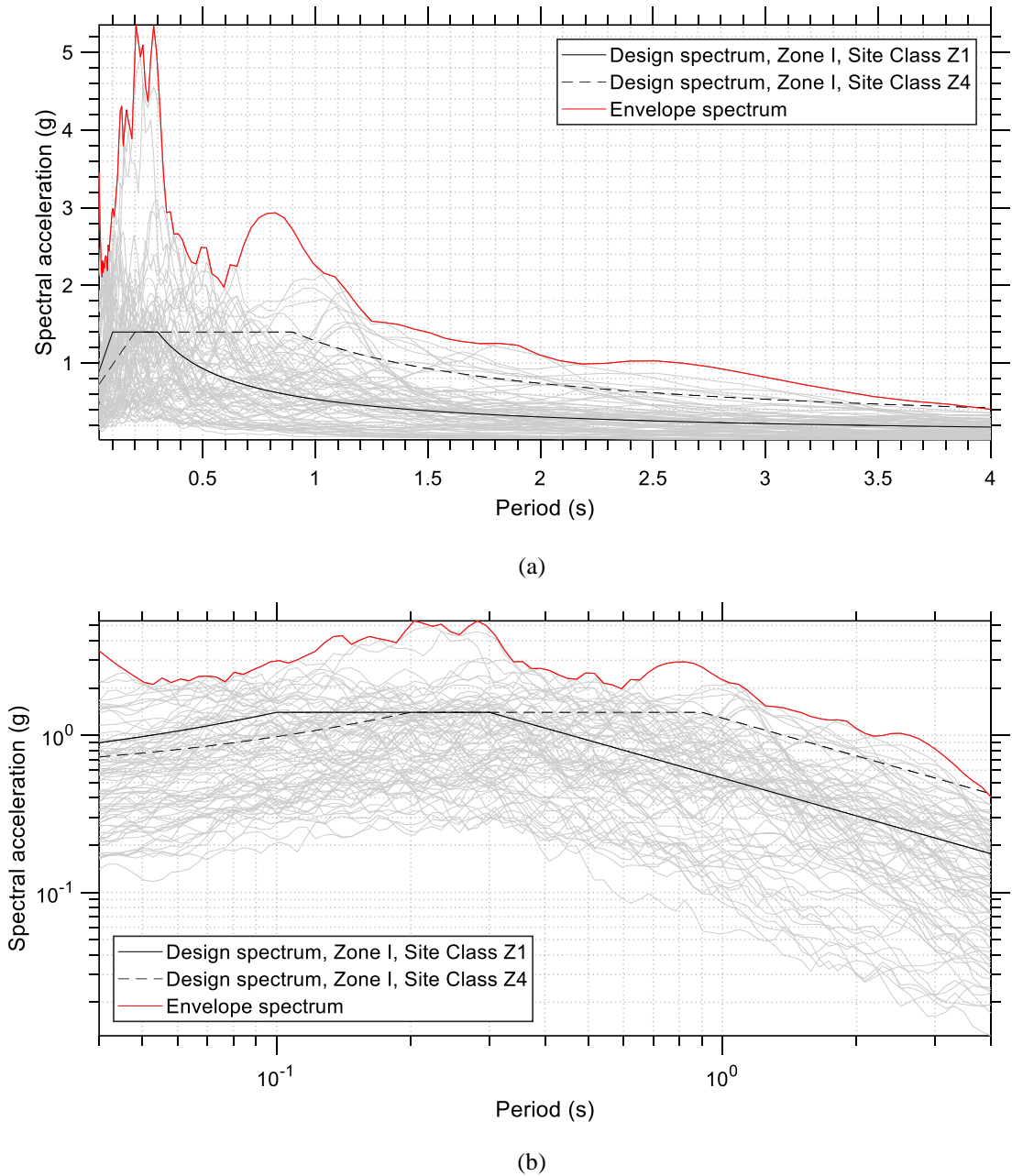

(a)

(b)

**Figure 18**. Design spectrum according to the Turkish Seismic code vs actual acceleration response spectra ($\zeta$=5%) of various records for the $M_w$ 7.8 earthquake: (a) Linear scale, (b) Logarithmic scale.

In Figure 18, 76 recordings have been taken into account in total, namely the two horizontal components (EW and NS) from the following 38 stations: TK_0201, TK_0213, TK_1213, TK_2308, TK_2708, TK_2715, TK_2718, TK_3115, TK_3117, TK_3123, TK_3124, TK_3125, TK_3126, TK_3129, TK_3131, TK_3132, TK_3133, TK_3134, TK_3135, TK_3136, TK_3137, TK_3138, TK_3139, TK_3141, TK_3142, TK_3143, TK_3145, TK_3146, TK_4617, TK_2703, TK_2712, TK_4615, TK_4616, TK_4624, TK_4629, TK_4630, TK_4632, TU_NAR.

## 3 Structural incremental dynamic analysis

The effect of an earthquake on structures can be quantified in various forms, such as by using the various peak and cumulative seismic parameters, as well as the response spectra, that were described in the previous sections. However, the engineer is often interested in monitoring the peak or cumulative structural response due to a seismic record, while varying a suitable intensity measure which is taken by appropriately scaling an earthquake record. This procedure is called Incremental Dynamic Analysis (IDA) and it involves performing multiple nonlinear dynamic analyses of a structural model under a ground motion record scaled to several levels of seismic intensity. The scaling levels are appropriately selected to force the structure through the entire range of behavior, from elastic to inelastic [11]. OpenSeismoMatlab [5] is capable of performing IDA analysis for a single record and a SDOF structure. Such IDA curves contain useful information about a seismic record, from a structural point of view.

### 3.1 Spectral acceleration – ductility curves

In Figure 19 and Figure 20 the IDA curves of a SDOF system for the horizontal components of the records TK3137 and KO-KHMN, respectively, are shown. These curves plot the spectral acceleration as an intensity measure (IM), and the ductility demand of the structure as a damage measure (DM). It is shown in Figure 19 that for structures with low yielding deformation, there is a higher ductility demand to withstand the same spectral acceleration (if possible). This trend is also observed in Figure 20. In addition, stiffer structures (with lower eigenperiods) seem to be more capable of withstanding higher spectral acceleration, as engineering intuition dictates. An important observation is the possibility of the fact that a structure can withstand more than one spectral acceleration for a single value of ductility (for example for $T$=1.5 s and $u_y$=0.1 m in Figure 19(a), or $T$=1 s and $u_y$=0.1 m in Figure 19(b). This is a common observation for structures responding in the elastoplastic regime. A general trend of the IDA curves is that with increasing intensity measure (spectral acceleration), the damage measure generally increases, as well.

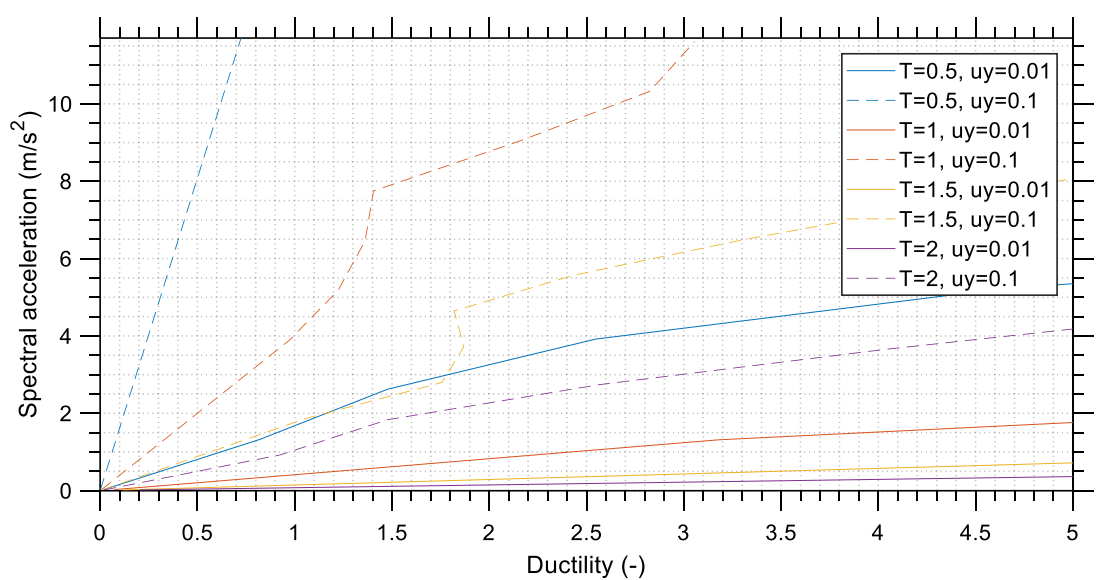

(a)

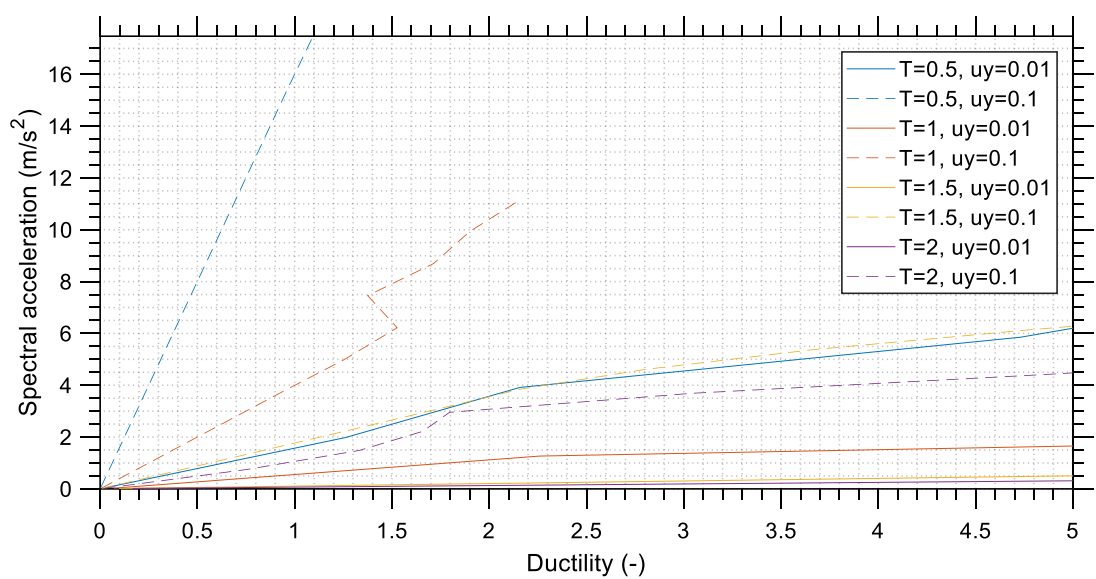

(b)

**Figure 19**. IDA - spectral acceleration vs ductility curves for the TK3137 record and for various combinations of yield displacement ($u_y$) and small strain eigenperiods ($T$) for the $M_w$ 7.8 earthquake: (a) EW component, (b) NS component.

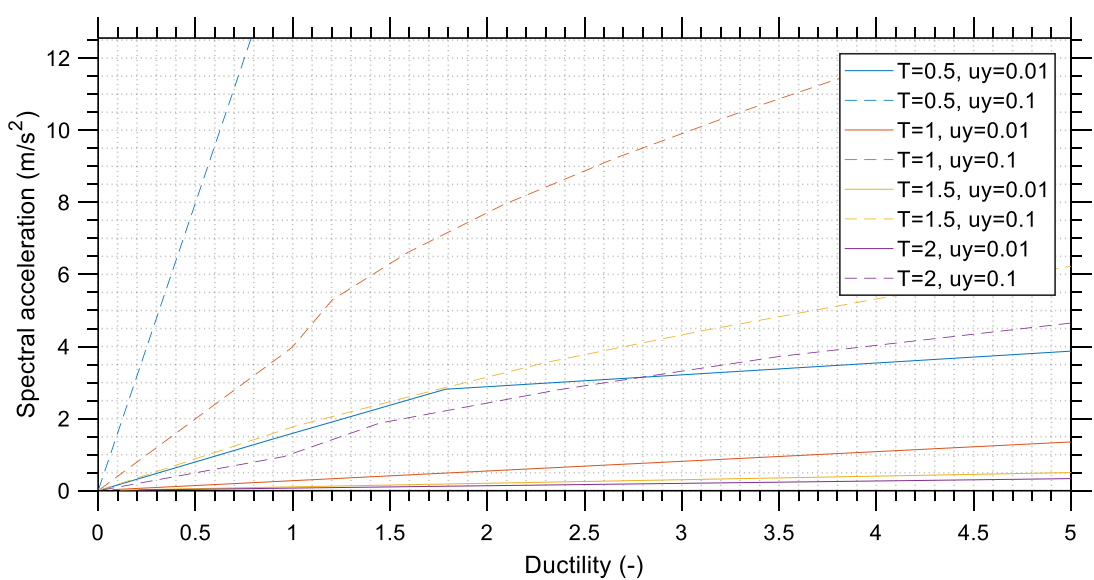

(a)

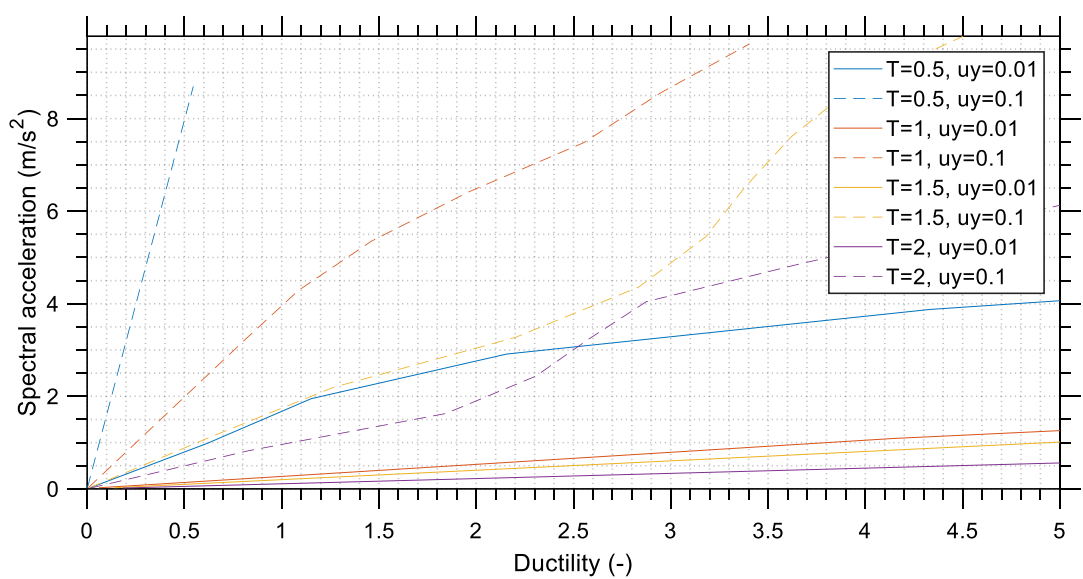

(b)

**Figure 20**. IDA - spectral acceleration vs ductility curves for the KO-KHMN record and for various combinations of yield displacement ($u_y$) and small strain eigenperiods ($T$) for the $M_w$ 7.8 earthquake: (a) EW component, (b) NS component.

## 3.2 Peak ground acceleration – ductility curves

In Figure 21 and Figure 22, IDA curves are presented regarding PGA versus ductility of a SDOF system for the horizontal components of the records TK3137 and KO-KHMN, respectively. The same trends as in section 3.1 are observed. Also, it is noted that in Figure 21(a), the curves for $T$=0.5 s and $u_y$=0.01 m and $T$=2 and $u_y$=0.1 m are nearly identical. This means that a stiff structure with low yield deformation can be equivalent to a flexible structure with moderate yield deformation. This has important implications for structural design. The latter type of structure is preferable since it is more economic. Therefore, reduced stiffness should be accompanied by moderate levels of yield deformation, to ensure that a structure will be able to withstand high earthquake acceleration levels.

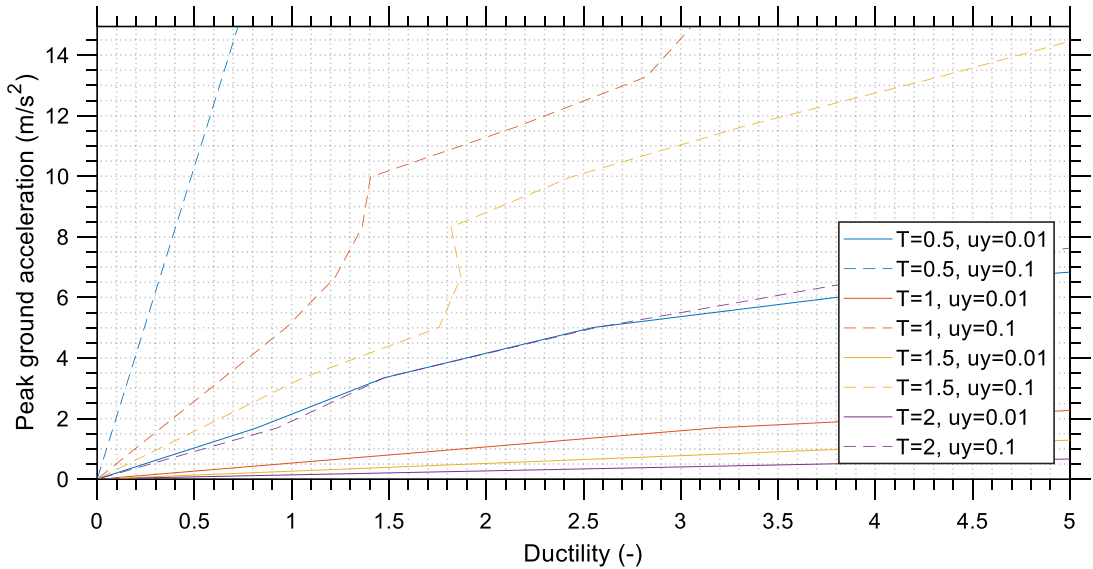

(a)

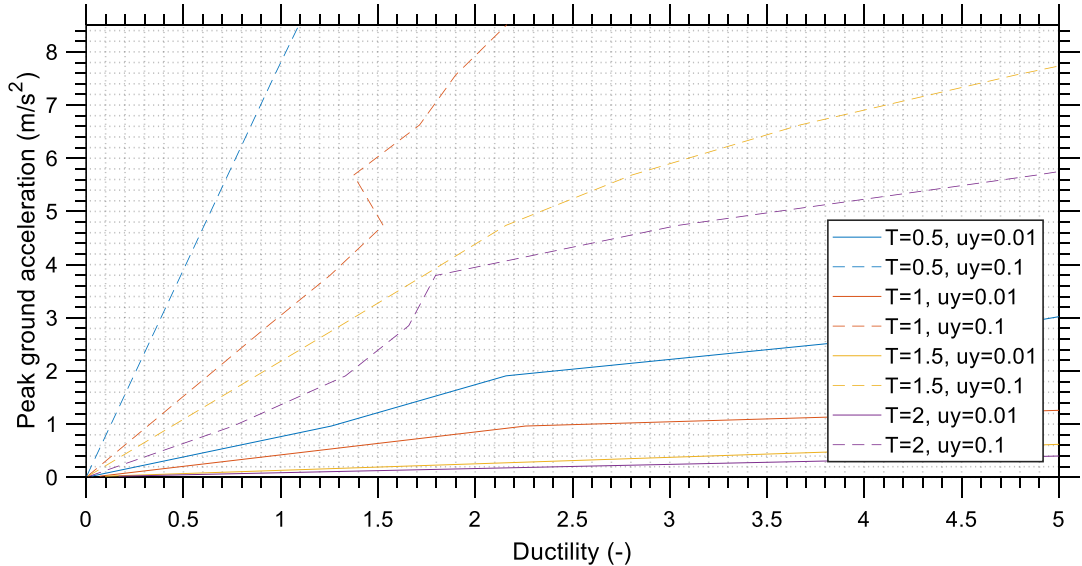

(b)

**Figure 21**. IDA - PGA vs ductility curves for the TK3137 record and for various combinations of yield displacement ($u_y$) and small strain eigenperiods ($T$) for the $M_w$ 7.8 earthquake:
(a) EW component, (b) NS component.

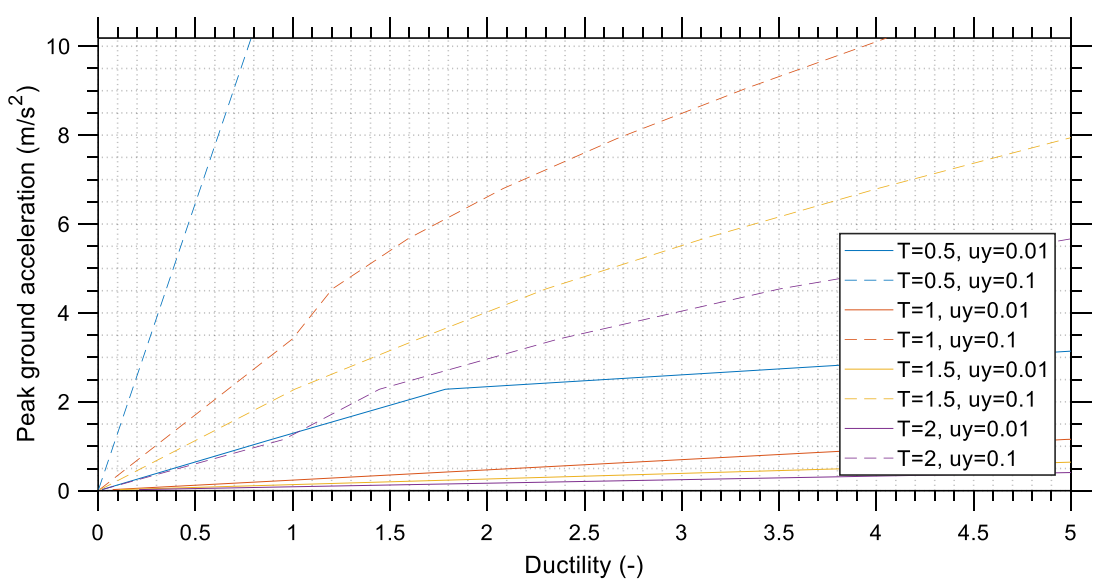

(a)

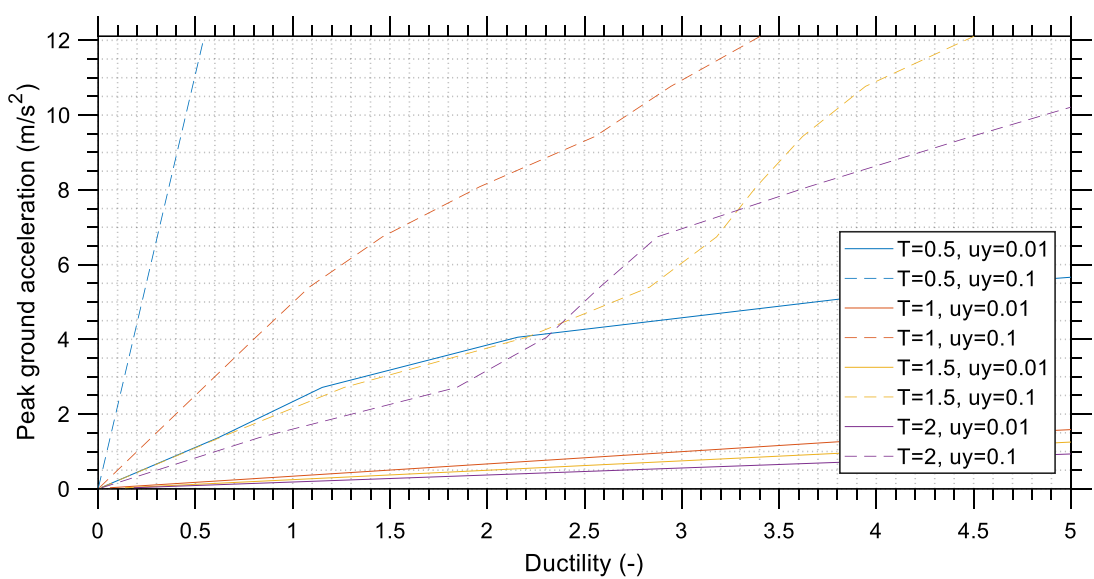

(b)

**Figure 22**. IDA - PGA vs ductility curves for the KO-KHMN record and for various combinations of yield displacement ($u_y$) and small strain eigenperiods ($T$) for the $M_w$ 7.8 earthquake: (a) EW component, (b) NS component.

## 4    Conclusions

The earthquake of magnitude $M_w$ 7.8 that hit Kahramanmaraş – Gaziantep regions in southern Türkiye on February 6, 2023, was a rare event of extremely large seismic power. This fact played a critical role in the intensity of the shaking that was experienced by structures and could provide some indirect hints explaining the large number of structural collapses. The acceleration spectral values of the seismic records were found to be substantially larger than the design acceleration spectrum values according to the Turkish seismic code. Moreover, this difference between the design and the actual response spectra covered a large interval of periods, which includes the eigenperiods of most buildings.

As expected, the maximum spectral acceleration of isoductile spectra is much lower than that of the corresponding linear elastic spectra. This implies a substantial difference in the seismic forces and shows the importance of structural ductility for proper seismic design. In many cases, the collapses due to the $M_w$ 7.8 earthquake were due to nonductile, brittle behavior, which in the case of reinforced concrete structures is closely related to under-reinforced structures. These structures, with very low ductility, responded in a more linear elastic – wise manner, and thus experienced much higher accelerations and forces.

A stiff structure with low yield deformation can be equivalent to a flexible structure with moderate yield deformation. This has important implications in structural design. The latter type of structure is preferable, since it is more economic. Therefore, reduced stiffness should be accompanied with moderate and not low levels of yield deformation, to ensure that a structure will be able to withstand high earthquake acceleration levels. Reduced design of under-reinforced concrete structures yielded a very low yield deformation and this could explain the large number of collapses due to the earthquake event.

Seismic codes worldwide (including the Turkish seismic code), have traditionally considered spectral acceleration as the most important parameter in structural design against earthquake loading. Nevertheless, the spectral velocity, which is usually ignored, appears to be an important parameter as well, which may exhibit a stronger relationship with the destructive effects of an earthquake and can therefore serve as a parameter from which to estimate the macroseismic intensity and structural damage.

### Conflict of Interest

The authors declare that the research was conducted in the absence of any commercial or financial relationships that could be construed as a potential conflict of interest.